\documentclass[11pt]{article}
\usepackage[margin=1in]{geometry}
\usepackage{amsmath,amssymb}
\usepackage{graphicx}
\usepackage{booktabs}
\usepackage[round,authoryear]{natbib}
\usepackage{caption}
\usepackage{subcaption}
\usepackage{xcolor}
\definecolor{linkblue}{RGB}{0,0,180}
\usepackage[colorlinks=true,citecolor=linkblue,linkcolor=linkblue,urlcolor=linkblue]{hyperref}
\usepackage{authblk}
\graphicspath{{figures/}}
\newcommand{\E}{\mathbb{E}}
\title{\bfseries Do In-Match Hydration Breaks Alter Match Momentum?\\
A Within-Match Case-Crossover Analysis of the 2026 FIFA World Cup}
\author[ ]{Debangan Dey\thanks{Corresponding author. Department of Statistics, Texas A\&M University,
College Station, TX, USA. E-mail: \texttt{debangan@tamu.edu}.}}
\affil[ ]{\small Department of Statistics, Texas A\&M University, College Station, TX, USA}
\date{July 2026}

\begin{document}
\maketitle

\noindent\textbf{Abstract:} The 2026 FIFA World Cup was the first to mandate hydration breaks, brief stoppages
near the twenty-second minute of each half, in every match regardless of weather. A widespread claim holds that
these breaks blunt the momentum of whichever side is dominant. We test the claim with a within-match
case-crossover design, using non-break minutes of the same match as self-matched controls. This differences out
every match-level characteristic and adjusts only for game time, scoreline, and pre-break momentum state. The
apparent post-break fade is not caused by the break: the dominant side's momentum decays at essentially the same
rate whether or not a break occurs, reverting through the stoppage as if play had continued. What the
break removes instead is the continuation of that side's pressure during the stopped minutes, and this leaves no
trace in net expected goals. The average effect is only weakly identified, resting on a trend extrapolation
into a window with no untreated minutes, whereas break-by-covariate interactions are stable; that average is
small and near zero. With only 99 matches the design lacks power to reject a small
effect, so we read the results by effect size and confidence interval. The estimates are not uniform across match states. The clearest modifier is team strength: the
break effect shifts by about $+1.6$ momentum points per $100$ Elo points of rating advantage, favouring a
stronger in-form side over a weaker side that is momentarily dominant, and it is the only
statistically significant interaction. A weaker pattern appears
by scoreline, where a comfortably-leading side of average strength shows an implied loss near $-4$ points on a
pre-break level of about $+20$. We report these state-dependent signals as exploratory and in need of
replication.

\bigskip

\section{Introduction}

For the first time, the 2026 FIFA World Cup made in-match hydration breaks universal: a stoppage of roughly
three minutes was administered near the $22^{\text{nd}}$ minute of each half in every match, regardless of
conditions. The policy drew immediate and pointed criticism from coaches and pundits, much of it framed
explicitly in the language of momentum. Emma Hayes, the United States women's head coach and a television
pundit during the tournament, observed that a break ``is advantageous for the team losing momentum,'' adding
``that's why I call them momentum breaks'' \citep{hayes2026momentum}. The England head coach Thomas Tuchel
objected that the stoppage ``interrupts and changes the identity of a football match,'' dividing the game
``almost in four quarters'' \citep{tuchel2026identity}, while the United States men's head coach Mauricio
Pochettino said plainly that it serves to ``cut the rhythm of the game'' \citep{pochettino2026rhythm}. The
tactical claim embedded in these reactions is that a break disproportionately harms the team that has built up
pressure, allowing a defending side to recover its structure and dissipating the attacking side's momentum.
This paper asks whether that claim is supported by the within-tournament evidence, and delineates the boundary
of what the data can answer.

\subsection{Related literature}

Environmental heat stress is a first-order determinant of physical output in football.
\citet{nassis2015heat}, using 2014 World Cup data, showed that higher wet-bulb globe temperature (WBGT) was
associated with reduced distance covered, motivating in-match mitigation. The WBGT index, which integrates
temperature, humidity, radiant heat, and air movement, has a long occupational history and well-documented
limitations when transported into sport \citep{budd2008wbgt}. Consensus guidance for competing in the heat
\citep{racinais2015consensus} and football-specific mitigation frameworks \citep{gouttebarge2023hottips}
recommend scheduled cooling or hydration breaks above high WBGT thresholds, of which the 2014 World Cup
implementation is the canonical example. The physiological case for the intervention is well supported:
controlled work shows that even brief in-play cooling breaks reduce thermal strain during football in the heat
\citep{chalmers2019cooling}, and simulation studies confirm the efficacy of the specific FIFA cooling-break
heat policy \citep{brown2025fifa}. Whether such stoppages also perturb the flow of the contest is a separate
and far less studied question.

The most direct precedent is the mid-half drinks break introduced when football resumed behind closed doors
during the 2020 pandemic. Analyses of that restart documented measurable shifts in match dynamics, from game
actions and tempo \citep{garciaaliaga2022covid} to the near-disappearance of home advantage
\citep{almeida2021covid,fischer2021ghost,link2022covid}; the additional stoppage, however, was only one
element of a bundled change that also removed crowds and expanded substitutions, and was not itself isolated
as a treatment. More generally, the frequency and duration of stoppages in football have been quantified
\citep{siegle2012stoppages}, and the type of interruption is known to shape how much time is added and how the
match state evolves \citep{li2024additional}. The cleanest causal evidence on whether a deliberate stoppage
interrupts momentum comes from other sports: \citet{gibbs2022timeout} estimate the effect of an NBA timeout on
an opponent's scoring run, and \citet{denhartigh2018tabletennis} show experimentally how a time-out disrupts
the build-up of psychological momentum in table tennis. We nonetheless found no peer-reviewed work estimating
the causal effect of a drinks or cooling break on subsequent scoring or momentum in football, and existing
discussion of ``momentum breaks'' remains journalistic, which is the gap the present paper
addresses.\footnote{The operational parameters of the 2026 policy (the near-$22^{\prime}$ timing and the
approximately three-minute duration) are documented in FIFA and IFAB material rather than in the peer-reviewed
literature, and are cited to those primary sources.}

The construct at stake is psychological momentum, the perception of progression toward a goal that in turn
alters subsequent effort and performance \citep{vallerand1988momentum,gernigon2010momentum}, and its study in
football is now substantial. Elite players report that they perceive momentum as real and consequential
\citep{redwoodbrown2018perceptions}, and a last-minute equalizer in knock-out football has been shown to shift
the subsequent course of a match \citep{denhartigh2020equalizer}. A parallel and instructive line of work has
subjected widely-held ``folk momentum'' beliefs to quasi-experimental scrutiny, often deflating them: the
supposed advantage of scoring just before half-time is contested \citep{gauriot2018halftime,meier2020myth,baert2018halftime},
the notion that a two-goal lead is uniquely dangerous does not survive careful analysis \citep{kent2025twogoal},
and apparent ``success breeds success'' effects can be largely mechanical rather than psychological
\citep{gauriot2019success}, though physical correlates of scoring and conceding are detectable in running output
\citep{konefal2024running}. This skepticism has deep roots in the ``hot hand'' literature, which cautioned that
perceived streaks are frequently indistinguishable from chance \citep{gilovich1985hothand,bareli2006hothand}
before a selection-bias correction partially rehabilitated the effect \citep{miller2018hothand}; separating
genuinely psychological momentum from strategic responses to the score remains the central identification
challenge \citep{meier2020tennis}, one that in-game win-probability models must also confront
\citep{robberechts2021winprob}. Complementary work quantifies the quality of individual goal-scoring chances
through models of shot conversion \citep{rathke2017xg,eggels2016xg,debdey2019shot} and the broader valuation of
attacking actions \citep{decroos2019vaep}, with continuing attention to model performance and reliability
\citep{mead2023xg}; these underpin the expected-goals (xG) metric that we use as an event-based outcome below.

The design draws on the case-crossover method of \citet{maclure1991casecrossover}, in which each case serves
as its own control by comparing a hazard window to referent windows drawn from the same subject, thereby
eliminating all time-invariant confounding by construction. Because validity hinges on how referents are
chosen, we follow the methodological cautions of \citet{mittleman1995control} and
\citet{marshall1993casecrossover}, and in particular the literature on time-trend bias:
\citet{lumley2000bias} and \citet{janes2005referent} show that naive referent schemes can bias estimates in
the presence of exposure trends, with time-stratified selection the preferred and essentially bias-free
strategy \citep{mittleman2005settled}. The design is related to the self-controlled case series
\citep{farrington1995sccs,whitaker2006sccs} and to segmented and controlled interrupted time series methods
\citep{wagner2002segmented,lopezbernal2017its,lopezbernal2018controls}. We frame the estimand within the
potential-outcomes account of causation \citep{rubin1974causal,holland1986causal,hernan2020whatif}, making
explicit the exchangeability, positivity, and SUTVA assumptions required to interpret a within-match break
contrast causally. For estimation we exploit the equivalence between within-subject (fixed-effects) and
correlated-random-effects specifications \citep{mundlak1978pooling}, with cluster-robust and
generalized-estimating-equation inference respecting within-match clustering
\citep{liang1986gee,cameron2015cluster}; where the response is a trajectory around the break rather than a
scalar, functional regression provides a natural extension \citep{morris2015functional,goldsmith2011penalized}.

\subsection{The measured outcome and a first look at the data}

\begin{figure}[t]
\centering
\includegraphics[width=\textwidth]{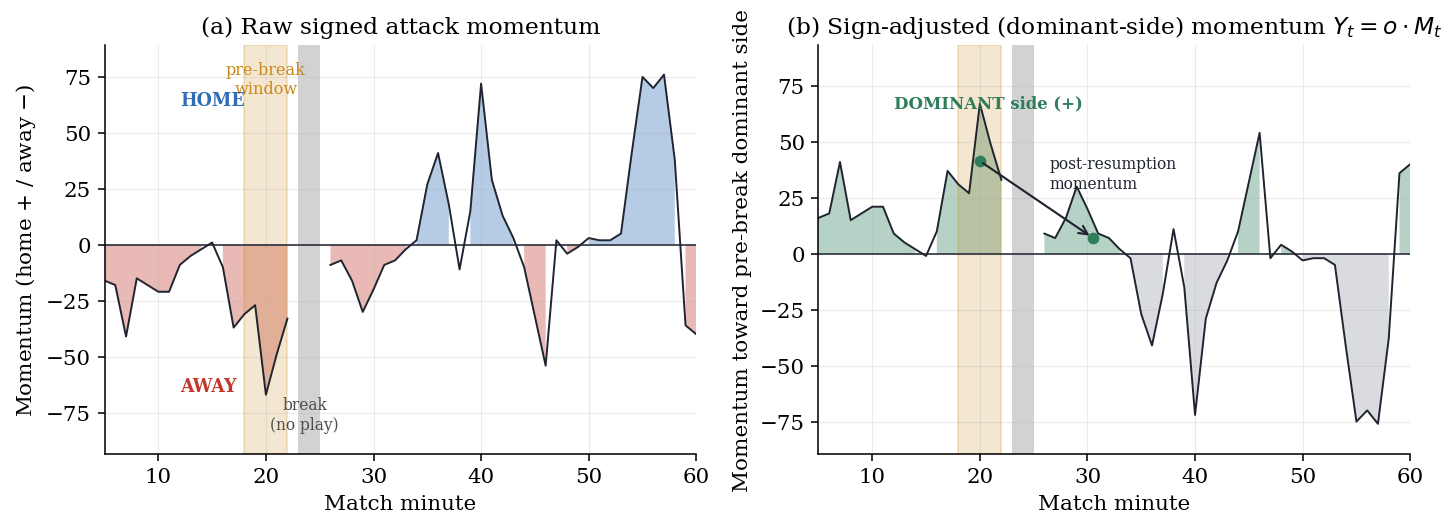}
\caption{\textbf{Constructing the sign-adjusted outcome, with the break shown as a stoppage.} (a) Raw signed
attack momentum for a representative match; the away side dominates the pre-break window (gold). The hatched
grey band is the approximately three-minute break, during which play is stopped and the momentum series
carries no informative value, and is therefore blanked. (b) After multiplication by the sign $o$ of the
pre-break mean, momentum is expressed toward the pre-break dominant side (positive by construction); the
effect of interest is the change from the pre-break level to the post-resumption level (arrow).}
\label{fig:orientation}
\end{figure}

Our outcome derives from Attack Momentum, a proprietary minute-resolved index published by SofaScore
\citep{sofascore2026} that
summarises, on a signed scale, which team is generating sustained attacking pressure. We orient it so that
positive values favour the home team and negative values the away team. Because the tactical claim concerns
the side that is dominant at the break, which is sometimes the home team and sometimes the away team, a raw
signed outcome conflates two questions. We therefore use a sign-adjusted (dominance-oriented) outcome. For a
break at match minute $c$ in match $i$, let
\begin{equation}
o_i \;=\; \operatorname{sign}\!\Big(\tfrac{1}{5}\textstyle\sum_{k=1}^{5} M_i(c-k)\Big) \in\{+1,-1\}
\end{equation}
be the sign of the mean raw momentum $M_i$ over the five minutes preceding the break. The oriented momentum is
$Y_i(t)=o_i\,M_i(t)$, so that a positive value always denotes momentum toward the pre-break dominant side
(Figure~\ref{fig:orientation}).

The Attack Momentum series is indexed by nominal match minute, and the match clock runs through a hydration
break, with the lost time added on later. Consequently the series carries ordinary-looking values during the
break, even though play is stopped; these are algorithmic decay artefacts rather than real play, and treating
them as post-break momentum would contaminate the outcome. We therefore exclude a uniform three-minute in-play
window at each break start (minutes $c,c{+}1,c{+}2$) and measure the post-break outcome only from resumption,
over $[c{+}3,\,c{+}12]$; the pre-break window is $[c{-}5,\,c{-}1]$. Observed break widths are tightly
concentrated (96\% are two or three minutes; Section~\ref{sec:data}), which justifies the uniform exclusion.
The same three-minute pseudo-break exclusion is applied to every control anchor, so that the treated-versus-control
contrast preserves an identical temporal structure.

A naive reading of Figure~\ref{fig:orientation}(b) suggests that the break was effective, since the dominant side is weaker afterward. Momentum is mean-reverting, however, and this before-and-after contrast is precisely what a sound design must resist: Section~\ref{sec:designs} shows that the dominant side's momentum reverts through the stoppage as if play had continued, decaying at essentially the same rate whether or not a break occurs, so the break neither creates nor prevents the fade. What it removes instead are the interpolated minutes of play in which the dominant side would have pressed its advantage, a reading we develop in Section~\ref{sec:discussion} alongside an event-based check that finds no effect on chance creation.

\subsection{Two features of the policy, and contributions}

\begin{figure}[t]
\centering
\includegraphics[width=0.62\textwidth]{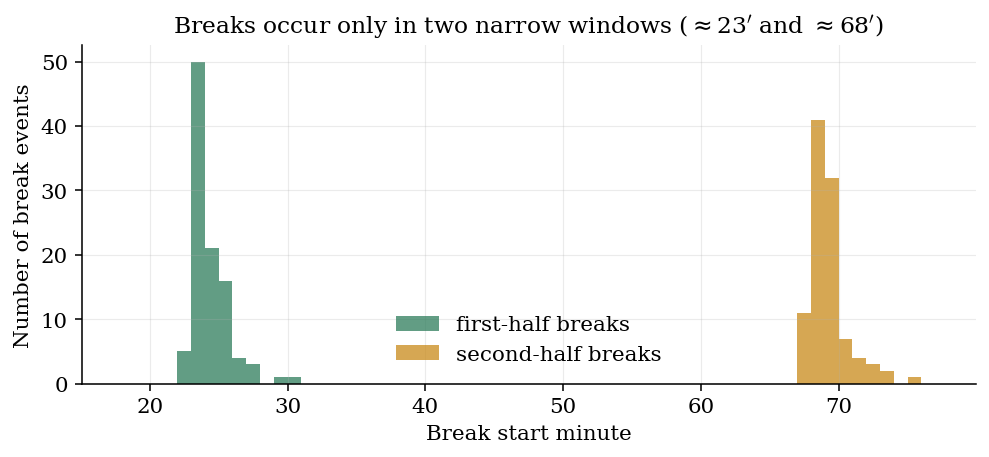}
\caption{\textbf{Breaks occur only in two narrow windows} ($\approx 23^\prime$, $\approx 68^\prime$). The
estimand is local to these minutes.}
\label{fig:minutes}
\end{figure}

Two structural facts shape the analysis. First, breaks occur only in two narrow windows: a first-half break
with median minute $23$ and a second-half break with median $68$ (Figure~\ref{fig:minutes}). As
Section~\ref{sec:ident} makes precise, the design identifies the effect of a break at the minute at which it
was administered, and cannot, in the absence of an untestable homogeneity assumption, speak to a break at
some other minute. Second, the policy breaks in every match, so that, unlike historical tournaments in which a
stoppage of this kind occurred only in hot conditions, heat and the administration of a break are not
separable in any across-match comparison. This entanglement motivates a purely within-match design, in which
such match-level factors are differenced out rather than modelled.

We make two contributions. First, we provide a design-based answer to a widely-asserted tactical claim: the
average break effect on the dominant side's momentum is small, but when read by effect size and confidence
interval rather than by statistical significance, the data remain compatible with competitively-relevant,
state-dependent effects, concentrated on leading teams, that we do not dismiss. Second, we characterise what a
case-crossover design can identify when treatment recurs at fixed times. The break main effect is recoverable
only by extrapolating a within-match time trend into a window that contains no untreated observations, and is
correspondingly fragile, whereas break-by-covariate interactions are identified from cross-sectional variation
among the treated and are stable.

The remainder of the paper is organised as follows. Section~\ref{sec:data} describes the data and the
cleaning steps. Section~\ref{sec:designs} sets out the two candidate designs for the break-window
counterfactual. Section~\ref{sec:ident} develops the within-match case-crossover, its identification and its
estimates, and Section~\ref{sec:external} the external-control robustness with its different estimand.
Section~\ref{sec:discussion} interprets the results and Section~\ref{sec:conclusion} discusses their broader
implications.

\section{Data}
\label{sec:data}

\subsection{Data collection}

For every 2026 World Cup match with a published Attack Momentum graph we retrieved the minute-resolved series
from SofaScore \citep{sofascore2026}, together with goals and their minutes. The series is a signed index (home positive, away
negative) posted at nominal-minute resolution: a regulation match yields approximately $92$ points (the $90$
nominal minutes plus a short trailing stub), with half time at the $45^{\text{th}}$, and an extra-time match
approximately $124$. The nominal indexing implies that the series does not carry a separate slot for every
second of first-half stoppage time, which the clock accrues during a break and the referee adds after the
$45^{\text{th}}$ minute. Because breaks occur at nominal minutes $\approx 23$ and $\approx 68$, well inside
regulation play, this feature does not affect their placement.

Exact break minutes were read from live match commentary rather than assumed at a fixed minute. For each match
we located the commentary markers that open and close the hydration break, yielding a start minute for each
break (and, for group-stage matches, an end minute). The resulting break minutes vary across matches around
the nominal $22^{\prime}$ and $67^{\prime}$ targets (Figure~\ref{fig:minutes}). Each event further carries the
current score margin and the pre-break momentum level and slope, all computed from the series, together with
match-dated international Elo for both teams, a WBGT heat index derived from venue coordinates and kickoff
time via a reanalysis weather service, and the local kickoff hour. Elo and WBGT are match-level constants;
they play no confounding role in the within-match design (Section~\ref{sec:ident}) but serve as effect
modifiers and for description.

We verified the minute alignment of the series in three ways. First, the hydration break leaves a visible
signature: momentum decays monotonically from the annotated break-start minute over the following few minutes,
consistent with a lull in which no play occurs, and this decay begins at the commentary-derived break minute,
confirming that break timing and the momentum index share the same axis. Second, we cross-validated against
goals: in about $80\%$ of first-half goals and $65\%$ of second-half goals the momentum in the minute of the
goal points toward the scoring side, with the remaining mismatches being against-the-run-of-play goals (a
counter-attack converted while the conceding side held the momentum) rather than an indexing offset. The
somewhat lower second-half agreement is consistent with the greater accumulation of stoppage time later in the
half, and does not affect the near-$23^\prime$ and near-$68^\prime$ break windows on which the analysis relies.
Third, we gated on series length, as described below.

\subsection{Cleaning}

Four cleaning steps were applied. First, two 2022-reference venues had transposed coordinates that returned
implausible temperatures; these were corrected before WBGT computation. Second, climate-controlled or domed
stadia (Atlanta, Houston, and Arlington, together with Las Vegas and Glendale) were assigned an indoor WBGT of
$20.7^\circ$C keyed on city name, since outdoor reanalysis does not reflect the indoor environment. Third, the
three in-play minutes at each break were blanked and the post-break outcome measured from resumption, as
described in Section~\ref{sec:ident}. Fourth, because the analysis relies on nominal-minute indexing, we
excluded any match whose momentum series length is inconsistent with that indexing. One match (Mexico versus
South Africa) returned an empty series, and two group-stage matches (Argentina versus Algeria, $136$ points,
and Austria versus Jordan, $132$ points) returned over-long series carrying interpolation and padding
artefacts that would displace the minute axis and the break location. These three matches were dropped.

One further match, France versus Iraq (group stage, match identifier \texttt{15186769}), had only a
first-half break recorded in commentary, its second half having featured no hydration break. Rather than
retain a match that contributes a single break, we drop it. The result is a balanced panel of $99$ matches
with exactly two breaks each, that is, $198$ break events, in which every retained match contributes one
first-half and one second-half break.

\subsection{Descriptive statistics}

Table~\ref{tab:desc} summarises the analytic sample: $99$ matches ($67$ group, $32$ knockout), $198$ break
events, and $3{,}139$ non-break control anchors. Break start minutes are tightly clustered (first half median
$23$, IQR $23$--$24$; second half median $68$, IQR $68$--$69$), and observed break widths are almost all two
to three minutes. Conditions span a wide range (WBGT $17.8$--$34.7^\circ$C), and at the break the pre-break
dominant side is most often level ($98$ of $198$) but frequently leading or trailing, which supplies the
cross-sectional scoreline variation that identifies the interaction terms.

\begin{table}[t]
\centering\small
\caption{\textbf{Descriptive statistics for the analytic sample} (99 matches, 198 break events).}
\label{tab:desc}
\begin{tabular}{ll}
\toprule
Quantity & Value \\
\midrule
Matches (group / knockout)              & 99 \ (67 / 32) \\
Break events (one per half)             & 198 \\
Non-break control anchors               & 3{,}139 \\
First-half break start minute           & median 23 (IQR 23--24) \\
Second-half break start minute          & median 68 (IQR 68--69) \\
Break width                             & median 3 min; 96\% are 2--3 min \\
WBGT heat index (AC venues corrected)   & mean 24.2$^\circ$C (SD 4.3), range 17.8--34.7 \\
Air-conditioned-venue matches           & 24 \\
Local kickoff hour                      & median 17:00 (range 12--22) \\
Pre-break dominant-side lead state      & behind 41; level 98; ahead by 1 40; ahead by $\geq$2 19 \\
Oriented Elo gap of dominant side       & mean $+111$ (SD 227) \\
Pre-break oriented momentum level       & mean $+20.4$ \\
\bottomrule
\end{tabular}
\end{table}

\section{Two designs for the break-window counterfactual}
\label{sec:designs}

\begin{figure}[t]
\centering
\includegraphics[width=0.92\textwidth]{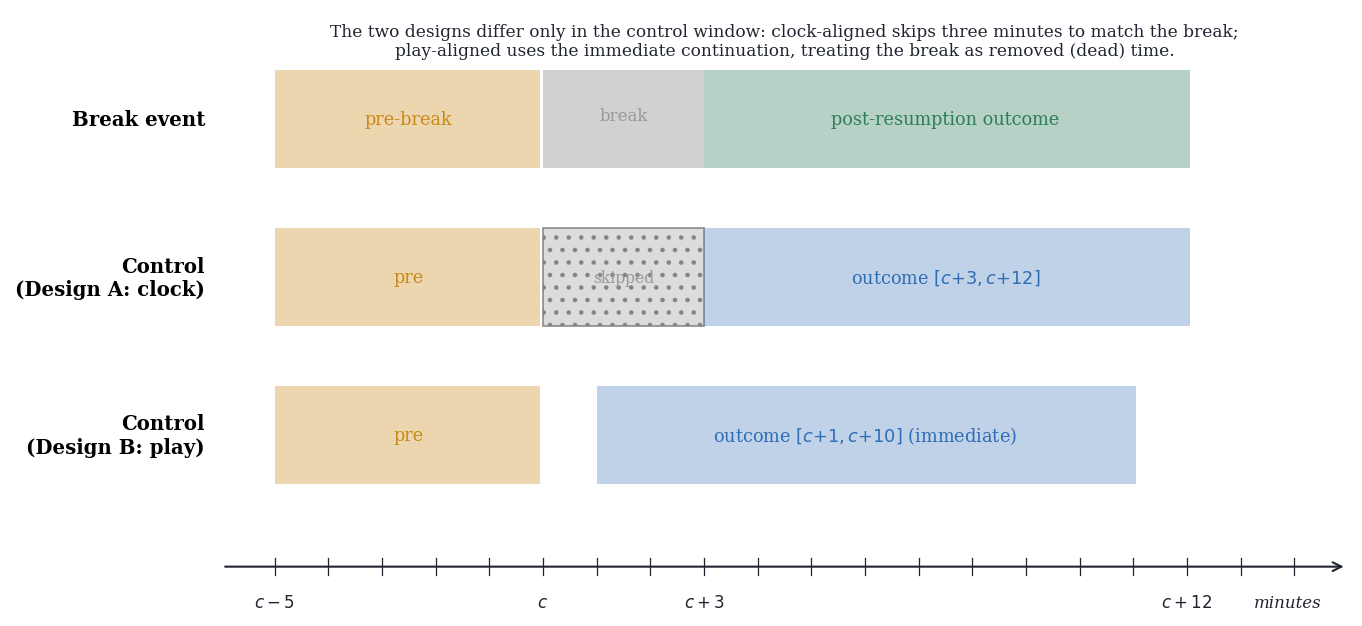}
\caption{\textbf{The two counterfactuals for the break window.} A break event's post-resumption outcome
window (green) is the same under both. Under the clock-aligned design a control skips three minutes to match
the break and its outcome begins at $c{+}3$; under the play-aligned design the control uses its immediate
continuation from $c{+}1$, treating the break as removed dead time. The two designs differ only in the
control window.}
\label{fig:designs}
\end{figure}

Identifying the break effect requires an assumption about what momentum would have done during the break
minutes, which are excluded from the outcome. Two natural counterfactuals are available, and we report the
analysis under both (Figure~\ref{fig:designs}). They differ only in how a break event's post-break window is
matched to a control's post-anchor window.

Under the clock-aligned counterfactual, momentum would have continued to evolve on the game clock during the
stoppage, so a break event and a control are compared at the same clock offset from their anchor. For a break
at start minute $s$ the outcome is the oriented mean momentum over $[s{+}3, s{+}12]$, and for a control at
minute $c$ over $[c{+}3, c{+}12]$, both beginning three minutes after the anchor. Under the play-aligned
counterfactual, the break is dead time during which momentum is suspended, so the resume minute is the first
minute of play and is matched to a control's first minute of play. The break event's outcome is then the
oriented mean over the ten play-minutes from its resume minute $e$ (the break end, taken per match), $[e, e{+}9]$, and
the control's is over its immediate continuation $[c{+}1, c{+}10]$. The break event's window is identical
under both counterfactuals; only the control window changes, advanced by the break duration under the
play-aligned assumption. Covariates and the game-time trend are evaluated at the anchor minute in both.

\begin{figure}[t]
\centering
\includegraphics[width=\textwidth]{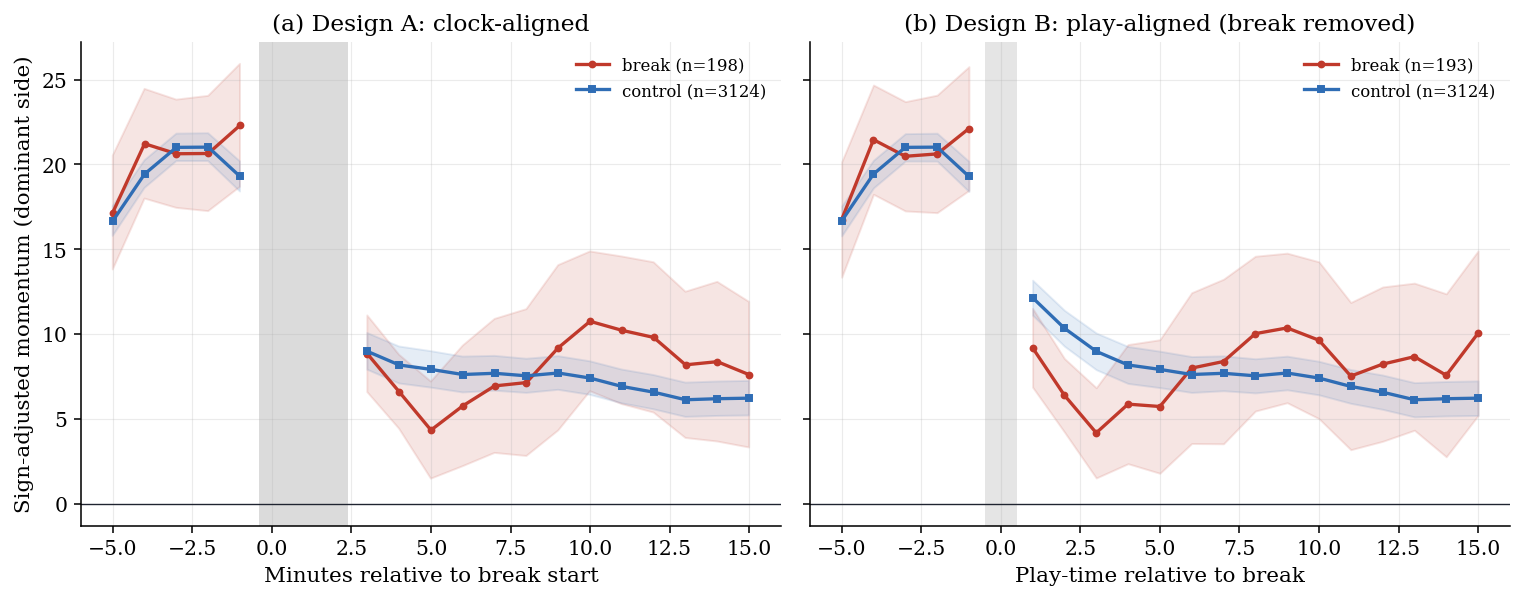}
\caption{\textbf{What the two designs show.} Event-time average of sign-adjusted momentum for break events
(red) and controls (blue), oriented toward the pre-break dominant side. (a) Clock-aligned: both series are
aligned on the game clock and the in-break minutes are blanked; the two decay alike, so the break neither
creates nor prevents the fade. (b) Play-aligned: the break is removed and the resume minute is play $+1$; the break events dip
below the continuous-play controls for the first few play-minutes, then converge. Bands are $\pm1.96$ pooled
standard errors.}
\label{fig:AB}
\end{figure}

The two designs give different pictures of the same data, shown side by side in Figure~\ref{fig:AB}. Under the
clock-aligned design (panel a), a break event and a control are compared at the same clock offset, and the two
decay in near-parallel from about $+20$ before the break to about $+8$ after it. The dominant side's momentum
reverts at the same rate whether or not a break occurs, so the break neither creates nor prevents the fade,
which unfolds through the stoppage as if play had continued. Under the play-aligned design (panel b), the break is removed and the
resume minute is treated as the first minute of play, so a break event is compared with a continuous-play
control at the same amount of play; here the break events sit below the controls for the first few play-minutes
before converging. That transient gap, present only under the play-aligned design, is confined to the window in
which a decaying momentum index under-reads after a stoppage, and Section~\ref{sec:discussion} shows with an
event-based outcome that it reflects the index rather than the team.

\section{The within-match case-crossover}
\label{sec:ident}

\subsection{The case-crossover target}

Index a candidate intervention point by $(i,c)$, denoting match $i$ and minute $c$. Let $D_{ic}\in\{0,1\}$
denote a break at $(i,c)$, and let $Y_{ic}(1),Y_{ic}(0)$ be the potential sign-adjusted post-resumption
momenta. In 2026 a break is administered at the two minutes $c\in\{c_{i1},c_{i2}\}$ near $23^\prime$ and
$68^\prime$, and all other within-match minutes are untreated. The target is the case-crossover average
treatment effect on the treated, local to the observed break minutes,
\begin{equation}
\tau \;=\; \E\big[\,Y_{ic}(1)-Y_{ic}(0)\;\big|\; D_{ic}=1\,\big].
\label{eq:estimand}
\end{equation}
The case-crossover design \citep{maclure1991casecrossover} takes each treated unit as its own control by
comparison to referent untreated occasions from the same subject, namely non-break minutes in the same match,
so that all stable subject-level characteristics cancel. The choice of referent minutes is the design's key
degree of freedom \citep{janes2005referent}.

\subsection{Identifying assumptions}

Write the within-match adjustment set as $X_{ic}=\{g(c),\,\text{half},\,L_{ic},\,S_{ic}\}$, where $g(c)$ is a
smooth game-time trend, $L_{ic}$ the pre-break momentum level and slope, and $S_{ic}$ the scoreline state.
Identification of \eqref{eq:estimand} rests on four assumptions. The first (A1) is conditional exchangeability
given $X_{ic}$ and a match effect $\alpha_i$: within a match, whether minute $c$ is a break minute is governed
by the clock and the referee's protocol rather than by the evolving state of play once game time and momentum
state are conditioned on. As a specificity check on this assumption, a placebo break placed at
$38^\prime/82^\prime$ returns no effect (Appendix~\ref{app:robust}), indicating that the estimator does not
manufacture effects at arbitrary non-break minutes. The second (A2) is no anticipation, in that pre-break play
is not a response to the upcoming break. The third (A3) is SUTVA, that
is, no cross-match interference and a single version of the treatment. The fourth (A4) is within-match
positivity, in that referent minutes with comparable $X_{ic}$ exist; this holds in the scoreline and
momentum-state dimensions but not in game time, since no untreated minute exists at the break location
(Section~\ref{sec:locality}). Under (A1)--(A4), the adjusted within-match contrast of post-resumption momentum
between break and referent minutes identifies $\tau$ \citep{rubin1974causal,hernan2020whatif}.

\subsection{Match-level covariates are absorbed by design}

A decisive feature of the self-matched design is that every characteristic fixed at the match level is
differenced out mechanically, whether or not we measure it. The two teams, their Elo ratings and Elo gap, the
WBGT, the kickoff hour, the stadium, and the tournament era are all constant across the break and referent
minutes of a match; they are collinear with $\alpha_i$ and cannot confound the within-match contrast. In
particular, Elo cannot enter as a confounding covariate: as a regressor alongside $\alpha_i$ it is perfectly
collinear and is dropped. Elo re-enters legitimately only as an effect modifier, $D_{ic}\times\text{Elo
gap}_i$, identified from cross-sectional Elo variation across matches, and we include it in
Section~\ref{sec:results}.

\subsection{The estimand is local to the break minute}
\label{sec:locality}

Because breaks occur only near $23^\prime$ and $68^\prime$ and we exclude the in-play window around each,
there is no untreated observation at the break location. The no-break momentum against which the main effect
is compared is therefore never observed; it is extrapolated from referent minutes elsewhere in the match via
$g(c)$ and the state controls. Two consequences follow.

The first is that the break main effect $\beta$ is a time-trend extrapolation and is therefore fragile. Since
$\beta$ leans on the shape of $g(c)$ carried into the unobserved window, it moves when $g$ is respecified or
when the referent minutes anchoring it are relocated (Table~\ref{tab:referent}, in \ref{app:tables}).
Polynomial trends leave
$\beta$ near zero, but saturated per-minute dummies, which absorb all game-time variation and leave nothing to
anchor the extrapolation, send $\beta$ to $-1.7$ with an interval of roughly $\pm 7$; relocating the referent
band moves $\beta$ from $+1.1$ to $-5.1$. The main effect is thus identified only under a smoothness
assumption on $g(c)$.

The second is that the interaction $\gamma$, by contrast, is contrast-identified and stable. The
break-by-scoreline interaction is the difference between how the outcome depends on the scoreline at break
minutes and how it depends on the scoreline at referent minutes, a contrast formed within each location from
cross-sectional variation in the scoreline among the break events. It does not rely on the time-trend
extrapolation, and accordingly it sits near $-0.2$ across trend specifications and, apart from the thin distal
``far'' referent band, is far more stable across referent windows than $\beta$ (Table~\ref{tab:referent}). In this sense the design can speak to how the break interacts with
match state even though it only weakly pins down the average level effect, and it cannot speak at all to a
break at an arbitrary minute without assuming homogeneity over game time, an assumption that is untestable
here and substantively implausible.

\subsection{The model}
\label{sec:model}

Our primary estimator is a within-match fixed-effects analysis of covariance with match-clustered standard
errors, fit on break and referent minutes jointly:
\begin{equation}
Y_{ic} = \alpha_i + g(c) + \phi\,\text{half}_{c} + \lambda L^{\text{lev}}_{ic} + \psi L^{\text{slp}}_{ic}
+ \delta S_{ic} + \beta D_{ic} + \gamma\,(D_{ic}\!\times\! S_{ic}) + \boldsymbol{\eta}^\top(D_{ic}\!\times\! Z_i)
+ \varepsilon_{ic},
\label{eq:model}
\end{equation}
where $\alpha_i$ is a match fixed effect; $g(c)=\theta_1 c+\theta_2 c^2$ (quadratic by default);
$L^{\text{lev}}_{ic}$ and $L^{\text{slp}}_{ic}$ are the oriented pre-break level and slope; $S_{ic}$ is the
oriented score margin; $D_{ic}$ is the break indicator; and $Z_i$ collects match-level modifiers (heat and Elo
gap) entered only in interaction. Standard errors are clustered on match $i$ \citep{liang1986gee,cameron2015cluster}.
Because the outcome is sign-adjusted, most between-match variation is removed and the residual intraclass
correlation is small and boundary-unstable; we therefore treat the fixed-effects estimator as primary and
report a random-intercept (Mundlak-style) mixed model \citep{mundlak1978pooling} as a corroborating check. The
match fixed effect $\alpha_i$ enforces the case-crossover logic exactly, and $g(c)$ makes break and referent
minutes comparable with respect to the natural evolution of momentum over the match (Figure~\ref{fig:AB});
under (A1)--(A4), $\beta$ together with the interaction terms is a plug-in estimate of $\tau$, subject to the
locality caveat of Section~\ref{sec:locality}.

\subsection{Estimates: a small average effect under both counterfactuals}
\label{sec:results}

We report and interpret effect sizes and their confidence intervals rather than declaring
results significant or null, following widely-endorsed guidance that a large $p$-value is not evidence of no
effect and that an interval should be read for the range of effects with which it is compatible
\citep{wasserstein2016asa,greenland2016misinterpret,amrhein2019retire}. This stance matters here for a
substantive reason: the policy is intended to be competitively neutral, so the quantity of interest is not
only the average effect but whether the break moves momentum in any match state by an amount that would be
material to the contest. For orientation, the pre-break dominant side sits at about $+20$ momentum points and
regresses naturally to about $+8$ over the outcome window (Figure~\ref{fig:AB}); an effect of a few points is
therefore a non-negligible fraction of the match's natural swing rather than statistical noise.

Table~\ref{tab:ladder} reports the full nested models under the clock-aligned counterfactual, and
Table~\ref{tab:coefB} (in \ref{app:tables}) the play-aligned counterfactual. The average break effect is small under both: the baseline break coefficient is
$+0.26$ (cluster-robust standard error $1.41$) under the clock-aligned design and $-0.63$ (standard error
$1.53$) under the play-aligned design. Both are centred near zero with intervals wide relative to the scale.
With only 99 matches, however, the design has limited power to detect a small average effect, so these wide
intervals are as much a statement about power as about the absence of an effect. The appropriate reading is
``no evidence of a large average effect'' rather than ``evidence of no effect,'' and the informative content is
the direction of the estimates rather than their statistical significance. The two designs differ only in the
sign of a small, imprecisely-estimated main effect and agree almost exactly on the covariate coefficients and,
as shown below, on the interaction structure. A random-intercept
mixed model returns nearly identical break-term estimates, consistent with a near-zero intraclass correlation.

\begin{table}[t]
\centering\small
\caption{\textbf{Nested within-match fixed-effects models, clock-aligned counterfactual} (outcome over
$[c{+}3,c{+}12]$ for both break events and controls). Coefficients with cluster-robust standard errors in
parentheses. The break $\times$ lead term is a single continuous slope on the oriented goal margin; the break
$\times$ Elo term is per $100$ Elo points. Match fixed effects included but not shown.}
\label{tab:ladder}
\begin{tabular}{lcccc}
\toprule
Term & Baseline & $+$ Break$\times$lead & $+$ Break$\times$Elo & Full \\
\midrule
Minute & $-0.09$ (0.16) & $-0.09$ (0.16) & $-0.09$ (0.16) & $-0.09$ (0.16) \\
Minute$^2$ ($\times10^{3}$) & $+0.85$ (1.27) & $+0.83$ (1.27) & $+0.84$ (1.27) & $+0.84$ (1.27) \\
Second half & $+2.41$ (4.21) & $+2.37$ (4.21) & $+2.39$ (4.20) & $+2.40$ (4.20) \\
Pre-break level & {\boldmath\bfseries $+0.24$ (0.04)} & {\boldmath\bfseries $+0.24$ (0.04)} & {\boldmath\bfseries $+0.23$ (0.04)} & {\boldmath\bfseries $+0.23$ (0.04)} \\
Pre-break slope & $+0.06$ (0.03) & $+0.06$ (0.03) & $+0.06$ (0.03) & $+0.06$ (0.03) \\
Score margin & $-2.31$ (1.20) & $-2.23$ (1.27) & $-2.24$ (1.26) & $-2.23$ (1.26) \\
Break & $+0.26$ (1.41) & $+0.38$ (1.49) & $+0.36$ (1.49) & $+0.35$ (1.49) \\
Break $\times$ lead & -- & $-0.90$ (1.39) & $-2.23$ (1.36) & $-2.19$ (1.37) \\
Break $\times$ Elo/100 & -- & -- & {\boldmath\bfseries $+1.59$ (0.80)} & {\boldmath\bfseries $+1.58$ (0.81)} \\
Break $\times$ heat/$^\circ$C & -- & -- & -- & $-0.21$ (0.31) \\
\midrule
Match fixed effects & Yes & Yes & Yes & Yes \\
Observations & 3337 & 3337 & 3337 & 3337 \\
Matches & 99 & 99 & 99 & 99 \\
\bottomrule
\end{tabular}
\end{table}

\subsection{Heterogeneity by lead and by heat}
\label{sec:within-cate}

\begin{figure}[t]
\centering
\includegraphics[width=\textwidth]{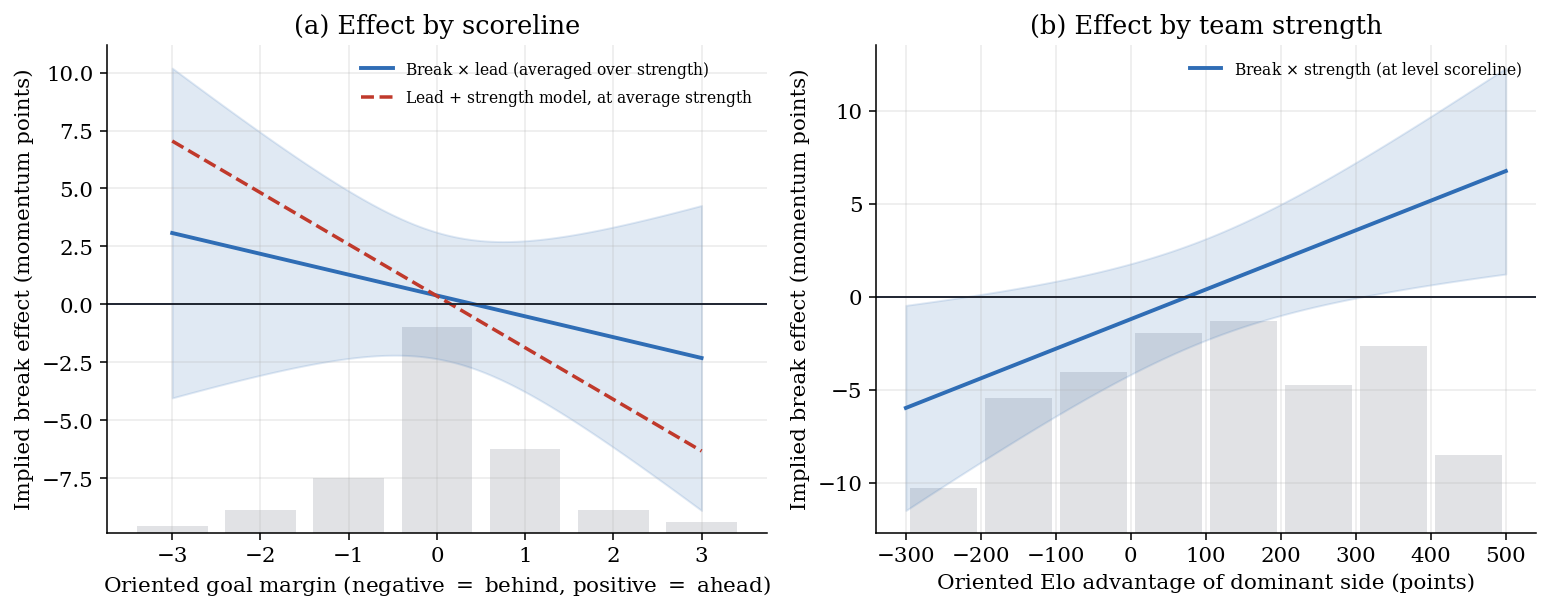}
\caption{\textbf{Implied break effect on the in-form side, by match state.} (a) As a function of the oriented
goal margin: the solid line is the continuous break $\times$ lead slope averaged over team strength; the dashed
line holds team strength at its mean. (b) As a function of the dominant side's oriented Elo advantage, at a
level scoreline. Shaded regions are $95\%$ confidence bands; grey bars show the distribution of break events
(data support). Negative values denote momentum moved against the pre-break dominant side.}
\label{fig:effect}
\end{figure}

The average masks systematic variation with match state, which is precisely the aspect on which a
competitive-neutrality concern should focus, and the interaction structure is nearly identical across the two
counterfactuals. The lead interaction is a single continuous slope on the oriented goal margin rather than a
set of categorical cells. The estimated break effect changes by about $-0.9$ momentum points per goal of lead
when averaged over team strength, and by about $-2.4$ per goal once team strength is held fixed, in both
counterfactuals and using all $198$ break events (Tables~\ref{tab:ladder} and \ref{tab:coefB}). The two slopes
differ because lead and strength are correlated: the side that is both dominant and ahead is
disproportionately the pre-match favourite, so each modifier masks the other until both are included. The
break $\times$ Elo term is positive (about $+1.6$ per $100$ Elo points; Table~\ref{tab:ladder}), so a stronger
in-form side sustains momentum through the break better than a weaker side that is momentarily dominant. This is
the only statistically significant interaction: at a level scoreline the
implied effect runs from about $-4$ momentum points for a side dominating despite a $200$-point rating deficit
to about $+5$ for a $400$-point favourite (Figure~\ref{fig:effect}b).

Figure~\ref{fig:effect} plots the implied break effect as a function of the oriented goal margin, with a
$95\%$ confidence band. At a level scoreline the effect is essentially zero ($+0.4$); it is mildly positive
for a trailing in-form side and negative for a leading one, reaching an implied $-1.4$ averaged over strength,
or $-4.1$ at average strength, at a two-goal lead. The band widens toward large margins because few break
events occur there, so the effect in that region is an interpolation along the fitted slope rather than an
independent estimate. At a two-goal lead the strength-adjusted interval is approximately $[-9.3,+1.1]$, which
is asymmetric about zero: it extends to $-9.3$ on the harm side but only to $+1.1$ on the benefit side, so the
data are considerably more compatible with the break eroding a leading side's momentum than with enhancing it.

Heat is the second candidate modifier. Splitting the within-match sample at a wet-bulb globe temperature of
$28^\circ$C, the break effect in cool matches is near zero under both designs ($+0.5$ clock-aligned, $-0.3$
play-aligned), whereas in hot matches it is negative ($-0.96$, interval $[-6.6,+4.7]$, clock-aligned; $-2.80$,
interval $[-8.3,+2.7]$, play-aligned; $n=42$ hot break events). The hot-match estimate leans against the
dominant side, more so under the play-aligned counterfactual, but its interval spans zero. We return to this
in Section~\ref{sec:discussion}, where the same pattern appears, less precisely, in the external design.

The state-dependent structure is exploratory: the large-lead and hot cells are thin (a lead of two or more
occurs $19$ times, and $42$ break events are hot), the modifiers were selected after inspection, and no
multiplicity adjustment is applied. It is a hypothesis for replication under the same universal-break policy
rather than a confirmed effect. Robustness and falsification checks for the within-match design, comprising the
stability of the interaction across trend and referent choices, a placebo break at $38^\prime/82^\prime$, and
sensitivity to the width of the in-play exclusion, are reported in \ref{app:robust} and are consistent
with the conclusions above.

\section{Robustness: external controls}
\label{sec:external}

\subsection{A different estimand and its data}

The within-match design compares a break minute to other minutes of the same match. A complementary design
compares each 2026 break event to no-break minutes drawn from other matches entirely. The estimand it targets
is different: a between-match contrast that no longer differences out the teams, the venue, the heat, or the
era, and therefore requires those factors to be adjusted for rather than eliminated by design. Writing the
external adjustment set as $X^{\text{ext}}_{ic}=\{Z_i,\,L_{ic},\,S_{ic},\,\text{half}\}$, where the match-level
modifiers $Z_i$ (Elo, WBGT, and local kickoff hour) now enter directly because they are no longer differenced
away, the design targets the same treated contrast as \eqref{eq:estimand},
\begin{equation}
\tau_{\text{ext}} \;=\; \E\big[\,Y_{ic}(1)-Y_{ic}(0)\;\big|\;D_{ic}=1\,\big],
\label{eq:estimand-ext}
\end{equation}
but must recover the counterfactual $Y_{ic}(0)$ from no-break minutes in other matches rather than from the same
match. Under between-match conditional ignorability, $Y_{ic}(0)\perp D_{ic}\mid X^{\text{ext}}_{ic}$, and
positivity, $0<\Pr(D_{ic}=1\mid X^{\text{ext}}_{ic})<1$,
\begin{equation}
\tau_{\text{ext}} \;=\; \E\big[\,\E[\,Y_{ic}\mid D_{ic}=1,\,X^{\text{ext}}_{ic}\,]-\E[\,Y_{ic}\mid D_{ic}=0,\,X^{\text{ext}}_{ic}\,]\;\big|\;D_{ic}=1\,\big],
\label{eq:estimand-ext-id}
\end{equation}
which the tight-caliper matching of Section~\ref{sec:matching} estimates on the region of common support. The
positivity condition in \eqref{eq:estimand-ext-id} is precisely what fails for the extreme-heat, early-afternoon
events (Section~\ref{sec:positivity}). We assemble
no-break control minutes from historical tournaments that did not administer universal breaks, namely Copa
Am\'erica (2019, 2021, 2024), the UEFA European Championship (2020, 2024), the CONCACAF Gold Cup (2019, 2021,
2023, 2025), and the 2018 and 2022 World Cups, retaining minutes in the same windows in which 2026 breaks
occur and carrying the same covariates: match-dated Elo, WBGT (with the same air-conditioning correction),
local kickoff hour, score margin, and pre-break momentum level and slope. All quantities are oriented toward
the pre-break dominant side, exactly as in the within-match design.

\subsection{The positivity limitation}
\label{sec:positivity}

This design has an identification limit that the within-match design does not. Because the 2026 policy breaks
in every match whereas historical stoppages of this kind occurred only in hot conditions, the extreme-heat and
early-afternoon games the policy most plausibly targets have essentially no historical no-break counterfactual.
Figure~\ref{fig:positivity} in \ref{app:positivity} shows the overlap: the historical control pool
has no match above a WBGT of $32^\circ$C, above which a dozen 2026 break events lie, and $28\%$ of break events
kick off at or before $14$:$00$ against $2\%$ of controls. The extreme-condition regime is therefore a
positivity void, and any external-control estimate there rests on extrapolation. This is the central reason we
treat the external analysis as a robustness check rather than a primary result, and it motivates trimming to
common support.

\subsection{Effect estimates under tight-caliper matching}
\label{sec:matching}

We estimate the effect by tight-caliper propensity-score matching \citep{rosenbaum1983propensity,stuart2010matching},
which trims to common support rather than reweighting into the void: each break event is matched to its nearest
control within a caliper of $0.2$ standard deviations on the estimated propensity logit, following the caliper
width recommended by \citet{austin2011caliper}, and events without a match inside the caliper are dropped. Only three break events are dropped, all early-afternoon games, consistent with the positivity
diagnosis. On the matched support the average effect is small and not distinguishable from zero under both counterfactuals, and, as in the
within-match design, more negative under the play-aligned assumption: the matched effect is $-1.57$ (interval
$[-5.00,+1.87]$) clock-aligned and $-2.48$ (interval $[-5.97,+1.01]$) play-aligned. The direction agrees with
the within-match crossover under the play-aligned assumption, where the effect is also negative, though every
interval crosses zero.

\subsection{Conditional effects}
\label{sec:ext-cate}

The external design is under-powered for the lead heterogeneity: balancing thin lead cells against historical
controls with limited overlap yields wide and non-monotonic estimates (behind $+1.4$, level $-0.1$, ahead by
one $+4.6$, ahead by two or more $-2.2$, all with intervals several points wide), so the within-match break by
lead slope of Section~\ref{sec:within-cate} remains the better-identified estimate of that heterogeneity. The
heat contrast is the one place the external design adds information, and it points the same way as the
within-match split: under the play-aligned counterfactual the effect in cool matches is $+0.43$
(interval $[-2.4,+3.0]$) and in hot matches $-4.60$ (interval $[-9.6,+7.6]$); these external-design intervals
are nonparametric bootstrap percentile intervals and need not be symmetric about the point estimate. The
hot-match point estimate is sizable and negative, but its interval is enormous, which is the positivity
limitation made concrete: the hot-condition effect is suggestive but not identified from historical data.

\section{Interpretation of the results}
\label{sec:discussion}

\subsection{Momentum reverts through the break as if play had continued}

The most instructive single result is Figure~\ref{fig:AB}. Because the clock-aligned and play-aligned
designs differ only in the control window, the clock-aligned comparison implicitly asks whether momentum
freezes during the stoppage. It does not: the oriented momentum of a break event falls during the break
minutes at essentially the same rate as a continuous-play control, so the series behaves as though play had
continued through the stoppage. The substantive implication is at the resumption. The dominant side re-enters
the match having missed the interpolated minutes of play during which, in a continuous game, it would have
continued to press its advantage, and that stretch is unobservable by construction because no football is
played during it. Whether the lost continuation carries any competitive cost cannot be settled by the momentum
index alone, which merely decays toward zero absent new events.

An event-based outcome speaks to the question. Net expected goals (net xG), a shot-quality measure built from
models of shot conversion \citep{rathke2017xg,eggels2016xg,mead2023xg,debdey2019shot}, accrues only on shots and is a true zero during a
stoppage rather than a decaying value, so it is free of the rebuild lag that produces the transient dip in
Figure~\ref{fig:AB}. Measured under both counterfactuals with the same per-match resume minutes, the
break effect on net xG is essentially zero under both counterfactuals ($-0.001$, interval $[-0.051,+0.049]$,
clock-aligned; $+0.011$, interval $[-0.051,+0.073]$, play-aligned). The dominant side creates the same quality of
chances after the break as a continuous-play comparison, whether aligned on clock-time or on play-time. The two
readings should be held together: the momentum series behaves as though the break excised a stretch of the
dominant side's pressure, yet on chance creation that removal leaves no measurable trace. We therefore read the
break as neither creating nor preventing the fade, while noting that the interpolated minutes it removes, though
without a detectable effect on net xG, are precisely the ones we cannot observe.

\subsection{Reading the effect sizes across scenarios}

Table~\ref{tab:summary} collects the break main effect across designs and counterfactuals. Every interval
crosses zero, but they are not all equally informative. The within-match clock-aligned estimate is small and
near zero, with an interval of about $\pm 3$ momentum points; on a scale whose natural swing over the window is
about $12$ points, an average effect larger than a few points is not well supported. We are cautious, however,
not to read this as positive evidence of a null: with only 99 matches the design has limited power to detect a
small average effect, so an interval that includes zero reflects that limited power as much as an absent effect. The
play-aligned estimates lean negative in every design, and the tight-caliper matched estimate is the most
negative ($-2.5$); none is individually distinguishable from zero, but the consistent negative sign under the
play-aligned assumption is the more informative feature.

\begin{table}[t]
\centering\small
\caption{\textbf{Break main effect on the dominant side's momentum across designs and counterfactuals}
(coefficient with $95\%$ interval). All intervals include zero; the play-aligned column leans negative.}
\label{tab:summary}
\begin{tabular}{lcc}
\toprule
Design & Clock-aligned & Play-aligned \\
\midrule
Within-match case-crossover        & $+0.26\ [-2.50,+3.02]$ & $-0.63\ [-3.63,+2.37]$ \\
External, tight-caliper matched     & $-1.57\ [-5.00,+1.87]$ & $-2.48\ [-5.97,+1.01]$ \\
\bottomrule
\end{tabular}
\end{table}

The state-dependent estimates require the opposite reading. The break by lead slope and the hot-match contrast
have wide, and in the leading-side case asymmetric, intervals, so significance testing would discard them; yet
their effect sizes carry a story a competitively-neutral policy should not ignore. A comfortably-leading side
of average strength has an implied effect near $-4$ momentum points with an interval reaching $-9$ and barely
crossing zero, and hot matches lean negative in both the within-match ($-2.8$ play-aligned) and external
($-4.6$) designs. These are not findings in the confirmatory sense, but their consistent sign and appreciable
magnitude mark exactly the conditions, a large lead or extreme heat, in which a break could plausibly move a
match, and in which the extreme-heat case is furthermore unidentifiable from historical data.

\subsection{Bringing the two designs together}

The within-match and external designs agree on the substance. The average break effect is small and not distinguishable from zero under both, and
the play-aligned assumption produces a mild negative lean in both. The lead heterogeneity is cleanly estimated
only within match, where the effect declines with the size of the lead; the external design is too thin here to
add more than imprecise corroboration. The heat heterogeneity is where the two designs reinforce each other in direction,
both placing the hot-match effect against the dominant side, while differing in precision, and it is also where
the external design's positivity void forecloses identification. The coherent picture is of a policy with no
detectable average effect and no detectable effect on chance creation, but with a state-dependent tendency, concentrated
on leading teams and hot matches, that the data cannot rule out and, in the extreme-heat case, cannot resolve.

\section{Discussion}
\label{sec:conclusion}

The strong form of the popular claim, that in-match hydration breaks reliably blunt the momentum of whichever
side is dominant, finds little support in these data. Much of what appears to be a break-induced fade is
momentum reverting through the stoppage as if play had continued, equally at non-break minutes, so the break
neither creates nor prevents the fade. What it removes is the continuation of the dominant side's pressure
during the stopped minutes, and this removal leaves no trace in net xG. The average break effect is small once the
in-play stoppage is excluded and match-level factors are differenced out by the case-crossover design. We are careful,
however, not to read this as evidence that the break has no effect. With only 99 matches, and two breaks each,
this sample does not have the power to reject a small average effect, so our failure to detect one reflects
limited power as much as an absent effect; a wide confidence interval that includes zero cannot
establish that the effect is zero, only that a large effect is not well supported. Read by effect size and
confidence interval, the data continue to admit momentum shifts of a few points. More importantly for a
policy intended to be competitively neutral, the effect is not uniform. For a comfortably-leading,
average-strength side the implied effect is about $-4$ momentum points, with an interval that reaches $-9$ and
only marginally crosses zero, so that the data are considerably more compatible with the break eroding a leading
side's momentum than with its enhancing it. This is a state-dependent, competitively-relevant effect that cannot be
ruled out, and because the policy aims for no match to be affected, it warrants monitoring and replication
rather than dismissal. Methodologically, when an intervention recurs at fixed times its average level effect
is only weakly identified, resting on a trend extrapolation into a window that contains no untreated data,
whereas its interactions with match state are cleanly identified from cross-sectional variation. The
state-dependent signal, rather than the average, is therefore where both the statistical information and the
policy relevance reside. These conclusions hold under both counterfactuals for the break window, the
clock-aligned and the play-aligned, and on an event-based outcome, xG, that is invariant to that choice
(Section~\ref{sec:discussion}). The external-control design agrees on the small, indistinguishable-from-zero
average, but cannot identify the extreme-heat regime the policy most plausibly targets.

\bigskip
\noindent\textbf{Acknowledgments:} The author gratefully acknowledges SofaScore
\citep{sofascore2026} for the publicly displayed Attack Momentum and match data on which this study is based.\\
\noindent\textbf{Research ethics:} Not applicable.\\
\noindent\textbf{Informed consent:} Not applicable.\\
\noindent\textbf{Author contributions:} The author is solely responsible for the content of this manuscript
and has approved its submission.\\
\noindent\textbf{Use of Large Language Models, AI and Machine Learning Tools:} A large language model was used,
under the author's supervision, to assist in compiling the dataset and in writing the analysis code. The author
verified all data, analyses, and results.\\
\noindent\textbf{Conflict of interest:} The author states no conflict of interest.\\
\noindent\textbf{Research funding:} None declared.\\
\noindent\textbf{Data availability:} The analysis relies on SofaScore's Attack Momentum index and
expected-goals values \citep{sofascore2026}, which are SofaScore's proprietary model outputs and are not
redistributed here. The match metadata, the commentary-derived hydration-break timings, and the derived
analysis datasets, together with a script that regenerates the SofaScore-sourced series from SofaScore's
public interface, are openly available in the repository at \url{https://github.com/Ddey07/wc2026-hydration-momentum}.\\
\noindent\textbf{Code availability:} All code that reproduces every figure, table, and reported quantity,
including a single end-to-end computational notebook, is openly available under the MIT License in the same
repository, \url{https://github.com/Ddey07/wc2026-hydration-momentum}.

\bibliographystyle{plainnat}
\bibliography{references}

\appendix
\renewcommand{\thesection}{Appendix~\Alph{section}}
\section{Positivity of the external-control design}
\label{app:positivity}

\begin{figure}[h]
\centering
\includegraphics[width=0.72\textwidth]{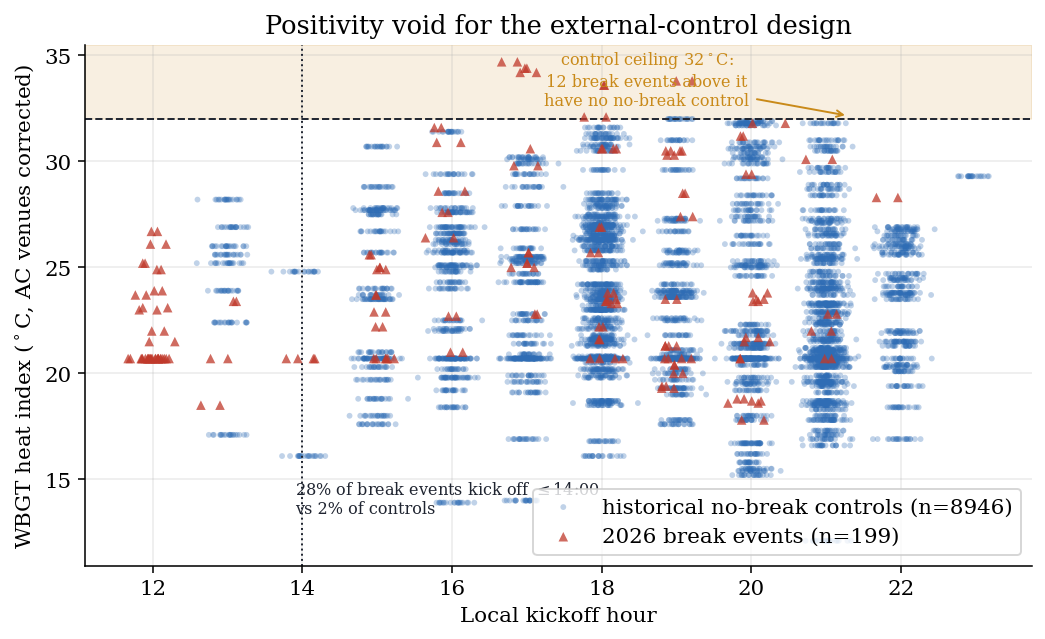}
\caption{\textbf{Positivity void for the external-control design.} 2026 break events (red) and historical
no-break control minutes (blue) by WBGT and local kickoff hour. The control pool has no match above the dashed
$32^\circ$C ceiling and few before $14$:$00$, so the hottest and earliest break events have no comparable
control.}
\label{fig:positivity}
\end{figure}

Figure~\ref{fig:positivity} displays the overlap between the 2026 break events and the historical no-break
control minutes in the heat-by-kickoff plane. The historical pool contains no match above a WBGT of
$32^\circ$C, above which twelve break events lie, and only $2\%$ of control minutes kick off at or before
$14$:$00$ against $28\%$ of break events. In precisely the extreme-heat, early-afternoon regime that the
mandatory-break policy most plausibly targets, there is no historical no-break counterfactual, so the
external-control estimate there is an extrapolation rather than an identified contrast.

\section{Supplementary tables}
\label{app:tables}

This appendix collects two tables referenced from the main text. Table~\ref{tab:coefB} reports the full nested
within-match models under the play-aligned counterfactual, the companion to the clock-aligned
Table~\ref{tab:ladder}. Table~\ref{tab:referent} reports the referent- and time-trend-sensitivity analysis that
supports the identification argument of Section~\ref{sec:locality}.

\begin{table}[h]
\centering\small
\caption{\textbf{Nested within-match fixed-effects models, play-aligned counterfactual} (outcome over ten
play-minutes from resumption for break events and over the immediate ten-minute continuation for controls,
using each match's actual break start and resume minutes). Cluster-robust standard errors in parentheses.
Compare with Table~\ref{tab:ladder}.}
\label{tab:coefB}
\begin{tabular}{lcccc}
\toprule
Term & Baseline & $+$ Break$\times$lead & $+$ Break$\times$Elo & Full \\
\midrule
Minute & $+0.03$ (0.16) & $+0.03$ (0.16) & $+0.03$ (0.16) & $+0.03$ (0.16) \\
Minute$^2$ ($\times10^{3}$) & $-0.60$ (1.18) & $-0.61$ (1.18) & $-0.61$ (1.17) & $-0.62$ (1.18) \\
Second half & $+2.31$ (4.05) & $+2.26$ (4.05) & $+2.28$ (4.03) & $+2.30$ (4.03) \\
Pre-break level & {\boldmath\bfseries $+0.23$ (0.04)} & {\boldmath\bfseries $+0.23$ (0.04)} & {\boldmath\bfseries $+0.22$ (0.04)} & {\boldmath\bfseries $+0.22$ (0.04)} \\
Pre-break slope & {\boldmath\bfseries $+0.16$ (0.04)} & {\boldmath\bfseries $+0.16$ (0.04)} & {\boldmath\bfseries $+0.16$ (0.04)} & {\boldmath\bfseries $+0.16$ (0.04)} \\
Score margin & $-1.98$ (1.09) & $-1.89$ (1.14) & $-1.89$ (1.14) & $-1.89$ (1.14) \\
Break & $-0.63$ (1.53) & $-0.49$ (1.59) & $-0.51$ (1.61) & $-0.53$ (1.60) \\
Break $\times$ lead & -- & $-1.04$ (1.36) & $-2.43$ (1.31) & $-2.37$ (1.32) \\
Break $\times$ Elo/100 & -- & -- & {\boldmath\bfseries $+1.66$ (0.82)} & {\boldmath\bfseries $+1.66$ (0.83)} \\
Break $\times$ heat/$^\circ$C & -- & -- & -- & $-0.30$ (0.31) \\
\midrule
Match fixed effects & Yes & Yes & Yes & Yes \\
Observations & 3355 & 3355 & 3355 & 3355 \\
Matches & 99 & 99 & 99 & 99 \\
\bottomrule
\end{tabular}
\end{table}

\begin{table}[h]
\centering\small
\caption{\textbf{The main effect is an extrapolation; the interaction is a contrast.} Gap-corrected
sign-adjusted momentum over $[c{+}3,c{+}12]$. Top: varying the game-time trend $g(c)$ with all referent
minutes. Bottom: fixing the break rows and relocating the referent band. The main effect $\beta$ swings; the
break$\times$lead interaction $\gamma$ is comparatively stable (the far band is a thin, distal slice).}
\label{tab:referent}
\begin{tabular}{lcc}
\toprule
Specification & Break main effect $\beta$ [95\% CI] & Break$\times$lead $\gamma$ [95\% CI] \\
\midrule
\multicolumn{3}{l}{\emph{Vary the time trend $g(c)$; all referent minutes}}\\
\quad linear                & $+0.26\ [-2.69,+3.21]$  & $-0.17\ [-2.77,+2.43]$ \\
\quad quadratic             & $+0.50\ [-2.42,+3.42]$  & $-0.15\ [-2.76,+2.46]$ \\
\quad cubic                 & $+0.92\ [-2.37,+4.21]$  & $-0.18\ [-2.79,+2.44]$ \\
\quad per-minute dummies    & $-1.70\ [-9.01,+5.61]$  & $-0.12\ [-2.65,+2.41]$ \\
\midrule
\multicolumn{3}{l}{\emph{Fix break rows; relocate referent band (H1 $\mid$ H2)}}\\
\quad adjacent $15$--$32\mid 60$--$77$ & $-0.31\ [-4.52,+3.90]$  & $+0.55\ [-1.98,+3.08]$ \\
\quad mid $11$--$40\mid 56$--$85$      & $+1.07\ [-2.68,+4.82]$  & $+0.48\ [-2.35,+3.30]$ \\
\quad far $6$--$15\mid 73$--$78$       & $-5.14\ [-13.46,+3.19]$ & $-2.34\ [-8.78,+4.10]$ \\
\quad all eligible                     & $+0.50\ [-2.42,+3.42]$  & $-0.15\ [-2.76,+2.46]$ \\
\bottomrule
\end{tabular}
\end{table}

\clearpage
\section{Robustness and falsification checks for the within-match design}
\label{app:robust}

The interaction estimates are stable across game-time trend specifications and, apart from the thin distal
``far'' referent band, far more stable across referent windows than the main effect (Table~\ref{tab:referent}),
consistent with contrast identification, whereas the break main effect is not, consistent with extrapolation. A placebo break at $38^\prime/82^\prime$ returns no effect, indicating that the
estimator does not manufacture effects at non-break minutes. The result is also insensitive to the width of
the in-play exclusion: widening it from three to four or five minutes leaves the average break effect small
and centred near zero ($+0.26$, $+0.64$, $+1.50$; all intervals wide and spanning zero) and the interaction
structure unchanged.

\end{document}